# How bio-friendly is the universe?

*P.C.W. Davies*
*Australian Centre for Astrobiology*
*Macquarie University, Sydney*

**ABSTRACT**

**The oft-repeated claim that life is 'written into' the laws of nature are examined and criticized. Arguments are given in favour of life spreading between near-neighbour planets in rocky impact ejecta (transpermia), but against panspermia, leading to the conclusion that if life is indeed found to be widespread in the universe, some form of life principle or biological determinism must be at work in the process of biogenesis. Criteria for what would constitute a credible life principle are elucidated. I argue that the key property of life is its information content, and speculate that the emergence of the requisite information-processing machinery might require quantum information theory for a satisfactory explanation. Some clues about how decoherence might be evaded are discussed. The implications of some of these ideas for 'fine tuning' are discussed.**

**NECESSARY VERSUS SUFFICIENT CONDITIONS FOR BIOPHILICITY**

There is now broad agreement among physicists and cosmologists that the universe is in several respects 'fine-tuned' for life. This claim is made on the basis that existence of vital substances such as carbon, and the properties of objects such as stable long-lived stars, depend rather sensitively on the values of certain physical parameters, and on the cosmological initial conditions. The analysis usually does not extend to more than these broad-brush considerations – that the observed universe is a 'well-found laboratory' in which the great experiment called life has been successfully carried out (Barrow and Tipler, 198?). So the conclusion is not so much that the universe is fine-tuned for life; rather, it is fine-tuned for the essential building blocks and environments that life requires. Such fine-tuning is a necessary, but by no means sufficient, condition for biogenesis. Thus 'anthropic' reasoning fails to distinguish between minimally biophilic universes, in which life is permitted but is only marginally possible, and optimally biophilic universes in which life flourishes because biogensis occurs frequently, i.e. life forms from scratch repeatedly and easily.

Rees has distinguished minimal from optimal biophilicity in relation to the known laws of physics (Rees, 2001). For example, if the value of the cosmological constant in our region of the universe is a frozen accident, and the probability of any particular value is uniformly distributed in the physically allowed range (this being >> the life-permitting range), then we might expect the actually observed value in our region to be not far from the threshold value at which life is permitted. If the cosmological constant were found to be, say, one million times smaller than the maximum permitted value, that would be evidence against anthropic selection. But a second issue is that, given the *necessary* condition that the known laws of physics lie within the 'well-found laboratory' range, is that *sufficient* for life to form with probability ~1 within, say, a Hubble volume? We could imagine a universe in which carbon and stable stars are abundant, but in which the emergence of life nevertheless required more. For example, it might require exceptional, fluky, physical conditions (such as the chance formation of some extremely unusual molecules). Alternatively it might require additional, yet-to-be-elucidated, laws or principles, possibly themselves requiring an element of fine-tuning. Following Shapiro (1986), I shall refer to this *second* distinct aspect of biophilicity as *biological determinism*. It is the assertion that life will be almost inevitable given earth-like conditions. Many astrobiologists are either witting or unwitting biological determinists. Some, such as de Duve (1995), believe that biological determinism is built into normal chemistry, others, such as Kauffman (1995), seek its origin in additional physical principles.

Conflation of necessary and sufficient conditions is common in discussions of astrobiology. For example, it is often claimed that because the stuff of life (C,H,N,O,P,S, and some organic molecules) are common substances, widespread in the universe, so too will life be widespread in the universe. But this is just as fallacious as claiming that because silicon is a cosmically abundant element so laptop computers will be widespread in the universe.

Another example concerns the existence of liquid water beyond Earth (e.g. on Europa) is often cited as a good reason to expect life there, on the basis that on Earth life is found almost everywhere that liquid water exists. One might indeed have legitimate reasons for doubting that life exists where liquid water is absent (e.g. on the Moon). Certainly liquid water is necessary for life as we know it. However, it is by no means sufficient. On Earth, aqueous habitats are invariably inhabited because the biosphere forms a contiguous system: life *invades* niches with liquid water, it does not emerge there *de novo*. So whilst it makes sense to follow the water when looking for extraterrestrial life, the mere existence of liquid water as such does little to raise expectations that life will actually be found.

Similarly, the abundance of carbon and the ubiquity of long-lived stable stars imply a bio-friendly environment, but on their own they do *not* imply that life will actually form. To draw that stronger conclusion involves an additional assumption: biological determinism. It might be in the form of a 'life principle' or, to use de Duve's evocative term, a 'cosmic imperative.'

Stated informally, a life principle might go something like this:

Consider a homogeneous medium of pre-biotic building blocks such as nucleotides and amino acids. Let the probability of assembling the simplest living organism solely from random rearrangements of the building blocks in unit mass of this medium in a duration $t \sim t_{universe}$ be $P_1$. Let the actual probability for life to emerge in this mass of medium be $P_2$.

Then the existence of a life principle implies $P_2 \gg P_1$. We may define the *amplification factor* as $P_2/P_1$.

As is well known, random molecular shuffling alone is exceedingly unlikely to make even a simple microbe from a planet covered in primordial soup, in the age of the universe. Hoyle (1983) has estimated $P_1$ at $\ll 10^{-40,000}$. By contrast, SETI proponents, who tacitly assume a life principle, have frequently asserted $P_2 \sim 1$ for a single earth-like planet, implying an enormous amplification factor of $\gg 10^{40,000}$.

What might be the cause of such stupendous amplification? Two popular theories are (i) molecular evolution, and (ii) self-organization. Theory (i) is really a re-definition of life. It asserts the existence of replication, variation and selection among a class of molecules of molecular weight $\ll$ the molecular weight of the simplest known living cell (Küppers, 1985). If small enough replicator molecules exist, they may form by chance in a suitable medium of modest mass with a probability $\sim 1$. An important unanswered question is then how fine-tuned the laws of physics need to be to permit the existence and replicative efficacy of these hypothetical molecules. Since we do not know what these molecules are, or even whether they exist, further progress on this matter must await future developments. It is possible to imagine, however, that the laws of physics would have to be even more stringently fine-tuned for such molecules to work as efficient Darwinian units. Theory (ii) is more easily studied, as several mechanisms of self-organization have been discussed in the literature (see, for example, Kauffman, 1995). However, as far as I know, there has been no study to determine how fine-tuned the efficacy of self-organization might be in relation to the laws of physics. It would be interesting to know, for example, whether elaborate and delicate metabolic cycles such as the citric acid cycle are sensitive to the mass of the electron or the value of the fine-structure constant.

Theories (i) and (ii) do not exhaust the possibilities for attaining a large amplification factor. There may exist (iii) new principles of complexity that will one day emerge from the general study of complex systems. I shall offer some speculations on option (iii) at the end of this paper.

So far I have dwelt on theoretical considerations. It is possible, however, that the matter of minimal versus optimal biophilicity in relation to biological determinism will be settled by observation. This would be the case if a second, independent, genesis of life were found on, say, Mars or Europa. Unless it could be demonstrated that our solar system as a whole offered exceptional conditions, it would then be reasonable to assert that life is widespread throughout the universe, and would arise with a high probability on most earth-like planets.

Before this conclusion is secure, however, we must confront the problem that there are two quite distinct ways in which life might be widespread in the universe. One is that the laws of nature and the cosmological initial conditions are such that life for emerges from non-life more or less automatically wherever there are earth-like conditions (this is the hypothesis of de Duve, 1995). The second is that life spreads efficiently across space – the so-called panspermia theory (Hoyle and Wickramasinghe, 1978). In the latter case, life may have started at just one location by an exceedingly improbable accident, but subsequently spread, establishing itself on a galactic or even cosmological scale during the multi-billion year history of the universe.

**STATUS OF PANSPERMIA AND TRANSPERMIA THEORIES**

The idea that life might be transported between planets across outer space was championed by Svante Arrhenius about a hundred years ago. Arrhenius (1908) envisaged microbes high in the atmosphere of a planet being propelled by the pressure of starlight until they reached velocities sufficient to escape from their planetary system altogether. If this were to happen in big enough numbers, there is a chance that a fraction of such expelled microbes might encounter a sterile but congenial planet elsewhere and 'seed' it with life. By implication, that is how life began on Earth, according to this theory. The panspermia theory makes no attempt to confront the problem of life's ultimate origin; it merely shunts it off to 'elsewhere.' There is no reason why panspermia cannot be combined with the assumption of multiple geneses of life, but the main attraction of the theory is that is would permit the universe to be teeming with life even if biogenesis was a unique event.

Panspermia has had few supporters in recent years, with the notable exception of Fred Hoyle and Chandra Wickramasinghe (1978). The main objection to the original theory is that the radiation environment of space is lethal to almost all known organisms. Hazards include solar and stellar ultra-violet, solar and stellar flares and cosmic radiation. Although examples of remarkable radiation resilience have been reported among certain terrestrial microbe species under special conditions (Minton, 1994), it remains true that all organisms would die quickly if exposed to direct solar ultra-violet, and more slowly (but still fast compared to interstellar transit times of millions of years) from cosmic radiation. It is possible to concoct elaborate scenarios in which microbes ejected from a planet are afforded a measure of protection from radiation (e.g. by coating in dust, immersion in an interstellar cloud or comet), enabling them to survive long enough to reach another star system, but such rare concatenations of events would not serve to provide a common dissemination mechanism to populate the galaxy, let alone to permit transits across intergalactic space. So whilst it may be the case that, here and there, one planet has seeded another in a neighbouring star system, a pervasive panspermia mechanism seems extremely implausible on current evidence.

The foregoing objections are largely circumvented, however, in a different scenario known as transpermia (Paine, 2002). In this theory, microbes are transported between planets cocooned inside rocks, which offer a measure of radiation and thermal protection (Davies, 1995, 1996, 1998; Melosh, 1997). Impacts by comets and asteroids with rocky planets are known to displace large masses of material. Theoretical studies by Mileikowsky et. al (2000) indicate that a substantial fraction of rocky ejecta would be displaced into orbit around the sun or parent star without suffering lethal shock heating. Some of these displaced rocks will eventually strike other planets and could thereby seed them with life. This is an old theory; its essential elements were articulated by Kelvin as long ago as 1871.

Transpermia would be a very efficient mechanism for transporting life between Mars and Earth, and to a lesser extend vice versa. Computations by Gladman *et. al.* (1996) show that 7.5 per cent of Mars ejecta will hit Earth eventually. Most microbes could withstand the *g* forces associated with impact ejection. The vacuum conditions and low temperatures of outer space need not prove lethal, as freeze-drying bacteria and archaea can actually increase their longevity. High-speed atmospheric entry would present the hazard of incineration, but rocks entering Earth's atmosphere at shallow angles would not invariably vaporise; fragments could reach the ground intact, and with short enough atmospheric transit times to prevent heat penetration to the interior. Mileikovsky *et. al.* (2000) have studied the radiation and thermal damage hazards to dormant bacteria and spores in this scenario, and determined that viability times of order millions of years are not unreasonable. This is easily long enough for live bacteria to make the journey from Mars to Earth.

The foregoing considerations make it almost inevitable that Mars and Earth will have cross-contaminated each other repeatedly during astronomical history. Mileikovsky *et. al.* (2000) estimate a traffic of about 4 billion tonnes of un-shocked martian material unheated above $100^{o}C$ reaching Earth over the last 4 Ga, and a smaller but significant amount going the other way. Given that Mars was warm and wet at a time when life is known to have existed on Earth, the seeding of Mars by terrestrial organisms seems very likely. The reverse is also true. In fact, a good case can be made that Mars was a more favourable planet than Earth for life to get started, raising the possibility that terrestrial life began on Mars, say 4.4. billion years ago, and spread to Earth subsequently (Davies, 1998; Nisbet and Sleep, 2001).

The key point about transpermia for the discussion in this paper is that it compromises the chances of finding a second genesis of life on our nearest neighbour planet. If traces of life are found on Mars it seems very likely that it would represent a branch of Earth life rather than an independent origin. The probability of contamination by Earth (and Mars) rocks diminishes sharply with distance, so there is a good chance that Europa is free of this problem. Transpermia would be a very inefficient mechanism to propagate life between star systems, as the probability that a rock ejected from Earth or Mars by an impact will hit an earth-like planet in another star system are negligible. The conclusion is that if biogenesis was a unique event, we might expect life to have spread beyond its point of origin to near-neighbour planets, but no further. If evidence for life were found outside the solar system (e.g. by detecting ozone in the atmospheres of extra-solar planets) it would provide strong support for biological determinism, with its implication of optimal biophilicity.

**THEORIES OF BIOLOGICAL DETERMINISM**

We have seen how life will not be widespread in the universe unless anthropic fine-tuning is augmented by the assumption of biological determinism. Attitudes to biological determinism fall into three categories:

A. It is false; life is a fluke restricted to Earth (or near neighbours).
B. It is true, and it follows as a consequence of known physics and chemistry.
C. It is true, but is not implied by known physics and chemistry alone; additional discoveries or principles are needed, perhaps to be found in the emerging sciences of complexity and information theory.

Position A was supported most notably by Monod (1971). Position B was adopted explicitly by Fox (Fox and Dose, 1977), who claimed evidence that the basic laws of physics and chemistry were biased in favour of generating biologically significant molecules. More recently de Duve (1995) has argued that whilst chemistry does not have 'life' etched into its principles in quite this manner, nevertheless biogenesis must be an expected product of chemistry. Position C has supporters in, for example, Eigen (Eigen and Schuster, 1979) and Kauffman (1995).

Evidence for B comes from pre-biotic chemistry, following the trailblazing experiment of Miller and Urey (Miller, 1953). The assumption was made by many that the Miller-Urey experiment was the first step on the road to life down which a chemical mixture would be inexorably conveyed by the passage of time. The common belief that 'more of the same' would eventually produce life from nonlife can be criticised as stemming from a nineteenth-century view of the living cell as some sort of 'magic matter' that can be cooked up in the laboratory by following an appropriate recipe; in other words, that biogenesis is primarily a problem of chemistry and chemical complexity.

An emerging view of life is that the cell is not so much magic matter as a supercomputer – a digital information processing and replicating system of enormous fidelity. Defining life through its informational properties rather than its chemical basis is akin to focusing on the software as opposed to the hardware. Obviously there are two aspects to biogenesis: the formation of an appropriate chemical substrate, and the emergence of an information-processing system. A fully satisfactory account of biogenesis requires an explanation for both hardware and software. So far, most of the research effort has been directed to the former. But according to the informational view of life, the nature of the hardware is secondary, since the essential information processing need not demand nucleic acids and proteins; it could be instantiated in alternative chemistry (Cairns-Smith, 1985).

According to the informational view, there are fundamental reasons why position B is deeply flawed. B theorists often remark that life is 'written into' the laws of physics. But this cannot be true as claimed. Life is a particular, very specific, state of matter. The laws of physics make no reference to specific states; they are completely general. The founding dualism of physics enshrines the independent status of eternal general laws and time-dependent contingent states. Any attempt to conceal states within the laws would introduce an element of teleology into physics, which is considered anathema by most scientists.

This argument can be sharpened by applying algorithmic information theory to the problem of life (Chaitin, 1990; Yockey, 1992). The laws of physics have very low information content: they describe how input information at time $t_1$ is converted to output information at time $t_2$, but they cannot add any information on the way. So the laws of physics cannot alone generate the informational content of life. Schrödinger clearly recognized this in his pioneering study of the structure of the genome (Schrödinger, 1944), which he termed 'an aperiodic crystal.' Normal, periodic, crystals are very low in information content. And crystals *are* written into the basic laws of physics: the structure of a crystal follows automatically from the geometrical symmetries encoded in those laws. But an information-rich molecule like a genome does *not* reflect the information content of the laws of physics, and so will not be generated inexorably from the operation of those laws. A genome shares with a crystal the property of stability, but the aperiodicity – or, more precisely, algorithmic randomness - is distinctive. Algorithmic information theory can make this argument rigorous by supplying a well-defined mathematical definition for the information content of a random sequence.

Molecular evolution is largely immune from the foregoing criticism, and falls under position C. In molecular evolution (or molecular Darwinism) the laws of physics and chemistry are augmented by the principle of natural selection, which enables information to be shunted from the environment into the cell. This may (or may not – we lack any proof) be sufficient to yield a form of biological determinism, especially if the phenomenon of convergent evolution was as manifest in molecular evolution as it is in normal Darwinian evolution (Conway Morris, 2003). Of course, this begs the question of why 'life as we know it' conveniently constitutes an attractor in the vast space of molecular complexity. It also raises the question of whether such canalised chemical pathways are sensitive to 'fine-tuning' of the laws of physics, and if so by how much.

Although molecular evolution might account for how information can accumulate in a molecular system once an information processing, replicating and storing mechanism exists, it fails to account for the origin of the information *processing* system itself. In other words, it offers a plausible account of the origin of the genetic database of early organisms, but not of the operating system at work in the cell. This is the same sort of distinction familiar from everyday computing: the database might be a list of addresses, for example, and the operating system Windows 2000. The database of addresses is useless without the Windows operating system to access and process it. In the same way, genetic information stored on a genome is no use on its own; it must be both interpreted and processed. Interpretation requires the operation of the genetic code, data processing requires a suite of proteins and other specialized molecules to implement the instructions in life's 'program.' It is far from clear that molecular evolution proceeding by purely Darwinian means of random variation and selection can create these key operating system features from scratch, even in principle.

**QUANTUM MECHANICS TO THE RESCUE**

I here offer some speculations that may cast light on a novel category C solution to biogenesis, viewed as a problem in information theory. First let me make a general point. Quantum information processing and quantum computation are now lively branches of physics (Milburn, 1988). The central message of these disciplines is that quantum mechanics provides a great - sometimes exponential - improvement in the information processing power of a physical system. If the problem of biogenesis is regarded as primarily a problem of information theory, then it is likely the theory of quantum information and quantum computation will cast important light on it. Obviously at some level quantum mechanics is important in life, but non-trivial quantum effects such as superposition, tunnelling and entanglement are normally considered to be incidental to the operation of life, owing to the problem of decoherence, which drives quantum systems very rapidly toward the classical regime when the system is immersed in a thermal environment (Zurek, 1982). There is, however, some circumstantial evidence that life may employ at least quantum-enhanced information processing, or even quantum computation as such (in limited contexts), by partially evading decoherence (McFadden, 2000).

How might quantum information processing help us to understand the origin of life? For the sake of illustration, suppose that the known tree of known life originated in something like the RNA world. This system is far too complex to arise *de novo*, so I further assume that there existed precursor replicator molecules simple enough to form by chance in a plausible prebiotic setting. The route from the latter to the former would have been long and tortuous. The first replicator would have faced a chemical decision tree of vast complexity, or which only one twiglet would represent the RNA world. The probability of the system following the right pathway from 'first replicator' to RNA world would be negligible if the chemical steps were conducted as a random walk. Therefore there must have been an amplification factor, as I have discussed. We can think of this either as a canalization of the reactions toward the RNA world, or of the RNA world acting as an attractor in the (vast) space of all molecular configurations. (These considerations would be invalidated if there were greatly many alternative routes to life, or a great many alternative forms of life. In this discussion I shall assume that life as we know it is more or less a unique possibility.) But there would be something suspiciously contrived and conspiratorial, not to say teleological, about the RNA world lying conveniently in a basin of attraction.

Is there another approach to this puzzle that avoids an element of implicit predestination? One could re-cast the concept of biogenesis in terms of a search problem: nature searches the chemical decision tree for a 'target' state – in this case the RNA world. But searching decision trees is one way that quantum mechanics can greatly improve efficiency. For example, Farhi and Gutmann (1998) have demonstrated an exponential improvement in search times for certain quantum decision trees.

Although there is no evidence that prebiotic soups employ quantum search techniques, there is a hint that quantum mechanics might enhance another type of search problem in biology: DNA replication. Patel (2002) has applied Grover's algorithm from the theory of quantum computation to the problem of the genetic code. Grover's algorithm would enable a quantum computer to search an unsorted database of N objects with a $\sqrt{N}$ improvement factor in efficiency. Patel finds that the numbers 3,4 and 20.2 emerge as solutions of the algorithm, numbers that he identifies with the triplet code, the number of nucleotides and the number of amino acids used by life, respectively. Patel has developed the outline of a theory (Patel, 2000) that goes beyond mere numerology and considers a model of DNA replication in which the nucleotides remain in a quantum superposition over a biochemically relevant timescale.

As remarked, the main argument against nontrivial quantum effects in biology is decoherence. In typical biochemical environments decoherence times are likely to be << reaction times. However, mechanisms are known that can greatly extend decoherence times. One such mechanism has been studied by Bell et. al. (2002) in the context of neutrino oscillations, but their model could also be applied to biochemistry. The basic idea is to consider a double potential well, L and R. A particle located in a given energy level in one well, say L, will tunnel back and forth between L and R. Now consider the entangled state $1/\sqrt{2}\,\{|L> + |R>\}$. If the system is weakly coupled to an external heat bath, representing a noisy environment, the entanglement will rapidly decohere, driving the system to a mixed state in which the particle is found in either L or R. But if the coupling to the environment is very strong, entanglement gets frozen in. What happens is that different energy levels of the system couple to each other via the environment, and this introduces a type of watchdog effect (McFadden, 2000) that freezes a region of the Hilbert space (a so-called decoherence-free subspace). Thus strong environmental coupling in effect ring-fences certain degrees of freedom, making them relatively immune from decoherence. One can imagine that this phenomenon creates quiet oases of Hilbert space in which quantum information may be processed over long enough timescales to create novel forms of molecular complexity.

These considerations amount to little more than pointers that nontrivial quantum information processing may take place in biologically relevant situations. What remains to be done is to identify how a quantum superposition recognizes a target state as biologically relevant, and can amplify its amplitude. Without this step, there remains an element of teleology, or predestination, in the way that chemistry 'discovers' life with unusual efficiency. Quantum mechanics improves over the classical formulation of the problem inasmuch as a quantum system may explore many pathways simultaneously, and 'collapse' onto a specific final state. But this factor alone is unlikely to solve the problem. It may have to be combined with some form of ratchet mechanism and a definition of the 'proximity to life' in the space of molecular configurations. Quantum ratchets are now under active investigation in both biological and nano-technological contexts (Linke, 2002).

If it transpires that quantum mechanics is indeed crucial to the emergence of life, it is likely that it will offer the most stringent application of fine-tuning. Weinberg (1993) has described quantum mechanics as logically isolated in the space of alternative theories, its status unique. Attempts to modify quantum mechanics, even very slightly, run into conflict with experiment. If quantum mechanics belongs to a set of measure zero in the space of theories, then life may turn out to be a truly singular phenomenon.